\documentclass[sigconf]{acmart}
\AtBeginDocument{%
  }

\setcopyright{acmlicensed}
\copyrightyear{2025}
\acmYear{2025}
\acmConference[IDC AI4CW '25]{Make sure to enter the correct
  conference title from your rights confirmation email}{June 23,
  2025}{Reykjavík, Iceland}

\begin{document}

\title{Challenges and Opportunities for Participatory Design of Conversational Agents for Young People's Wellbeing}

\author{Natalia I. Kucirkova}
\affiliation{%
  \institution{University of Stavanger}
  \city{Stavanger}
  \country{Norway}
}
\email{natalia.kucirkova@uis.no}

\author{Alexis Hiniker}
\affiliation{%
  \institution{University of Washington}
  \city{Seattle}
  \country{Washington}}
\email{alexisr@uw.edu}

\author{Megumi Ishikawa}
\affiliation{%
  \institution{Kokushikan University}
  \city{Tokyo}
  \country{Japan}
}
\email{megumi.i@kokushikan.ac.jp}

\author{Sho Tsuji}
\affiliation{%
 \institution{University of Tokyo}
 \city{Tokyo}
 \country{Japan}}
 \email{shotsuji@ircn.jp}

\author{Aayushi Dangol}
\affiliation{%
  \institution{University of Washington}
  \city{Seattle}
  \country{Washington}}
\email{adango@uw.edu}

\author{Robert Wolfe}
\affiliation{%
  \institution{University of Washington}
  \city{Seattle}
  \country{Washington}}
\email{rwolfe3@uw.edu}

\renewcommand{\shortauthors}{Kucirkova et al.}

\begin{abstract}
  This paper outlines the challenges and opportunities of research on conversational agents with children and young people across four countries, exploring the ways AI technologies can support children's well-being across social and cultural contexts.
\end{abstract}



\maketitle

\section{Introduction}

This conceptual paper examines the cross-cultural dilemmas encountered during research on conversational agents with children and young people across four different countries. We identify key challenges arising from cultural, ethical, and methodological differences, and propose pathways to foster interdisciplinary collaboration. In doing so, we explore how AI technologies can be designed to promote children’s digital well-being across diverse social and cultural contexts.

\section{About Conversational Agents}

As artificial intelligence (AI) continues to evolve, we are witnessing a shift from narrow AI applications to more general-purpose AI, including generative AI systems. These systems may incorporate a variety of background models, such as language models or image generation models. In our work, we focus specifically on the intersection between the backend, represented by language models, and the frontend, operationalized through so-called conversational agents (CAs for short). These AI-enabled tools, such as Alexa, Google Home, and similar systems, are capable of performing tasks such as playing music or coordinating shared activities, like managing a family calendar, in response to voice commands.

\section{Four research studies with CAs}

Our research team, supported by funding from the Jacobs Foundation and CIFAR, has conducted a series of studies over the past four years focusing on the use of CAs. We conducted one comparative study involving families in Japan, the US, and Norway, collecting data as part of a larger international project aimed at examining parents’ attitudes toward the use of CAs with their preschool-aged children \cite{kucirkova2024parents}. We also carried out a second comparative study in the US and Nepal to examine both: 1) how language technologies (specifically, large language models and static word embeddings) represent teenagers in both Nepali and in English, and 2) how teenagers in both Nepal and the US feel about these representations \cite{wolfe2024representation}.

Both projects were framed within the field of human–computer interaction, with a strong commitment to developing design recommendations that address both the benefits and limitations of current CA designs. In the first study, eleven Norwegian parents of children aged 3 to 5 years participated. Participants in the USA and Japan followed the same research protocol. In Japan, ten parents of 3 to 6 years of age participated. In our study of teens (aged 13-17) in the US and Nepal, we held synchronous or asynchronous workshops wherein teen participants completed data collection instruments while learning about the process of training AI via a narrative appropriate to the target age range. In the U.S., N=14 teen participants completed the workshop, predominantly respondents from our University’s Communication Studies Participant Pool; in Nepal, where we worked with educators residing in Kathmandu to recruit for the study, N=18 teen participants completed the workshop.

Working cross-culturally, we adhered to strict ethical guidelines and emphasized the importance of understanding cultural differences, including the need for geographically specific training foundations for conversational agents. Our findings revealed notable variations across contexts. For instance, Norwegian parents, unlike their US counterparts, identified several shortcomings in conversational agents when used with their children. In particular, they expressed concerns about the agents' limited support for children’s wellbeing and social skill development, noting the absence of polite language such as the use of please and thank you, and agents offering adequate support when children needed it. Moreover, parents highlighted the agents’ inability to recognize or accommodate local accents, which they felt undermined both their children's linguistic development and cultural identity \cite{kucirkova2024parents}.

In the study of US and Nepalese teens, we conducted studies of language technologies and teen participants largely in parallel, with teen participants responding to the same prompts that we presented to generative language models. This allowed us to observe the differences in 1) how teens and language models responded to the same prompt; and 2) how responses to prompts varied across languages and cultural contexts. We observed that the responses of language models to straightforward prompts like “At school, the teenager…” were significantly more sensationalistic than those of teen participants, with language models highlighting social problems (including issues like bullying, mental illness, and toxic technology use) in about 30\% of cases in English and in about 13\% of cases in Nepali. Teen participants, on the other hand, responded in ways that almost uniformly represented the positive aspects of teen life, and were mostly notable for just how mundane they were in comparison to AI; among the most “troubling” responses written by a teen participant is one that describes sleeping in class.    
Following our experience of designing these studies, carrying them out implementing in local contexts and reflecting on the findings, we have some noticed shared methodological challenges and opportunities for future studies with this type of AI.

\section{Challenges}

\begin{enumerate}
    \item \textbf{Ethics review variability:}
Ethical review processes vary significantly across countries, requiring different agreements, documentation, and approval mechanisms. Navigating these diverse requirements posed logistical and conceptual challenges when conducting cross-cultural research with AI.
\item \textbf{Translation and loss of meaning:}
Working with qualitative data demands careful attention to language precision. However, translating data across languages risks stripping away subtle meanings and cultural nuances. This challenge is further compounded by the limitations of current generative AI models, which do not sufficiently support nuanced, context-sensitive translation for research purposes.
\item \textbf{Youth participation:}
Involving children and young people is essential for the credibility and relevance of our research. This is particularly important given that the outputs of conversational agents often appear highly plausible, making it critical to correctly interpret children's reactions and insights. However, establishing meaningful connections with young participants requires time, trust-building, and a deep understanding of their contexts; it cannot be delegated to a conversational agent.
\item \textbf{Impact of Local Language Resources on Study Materials:}
In our study of US and Nepalese teens, we had to ask Nepalese teens to respond to our workshop questions on pen and paper. The reason for this was that, in order to make a direct comparison of teen responses with Nepali-language models, we needed to collect responses in Nepali, using Devnagari script. However, our Nepalese participants mostly did not have access to computing resources (i.e., keyboards and word processing programs) that would allow them to respond in their native language. Thus, research focusing on specific linguistic or cultural characteristics may need to adapt to limitations imposed by technological resources available in a low-resource or local language.
\end{enumerate}

\section{Opportunities}

\begin{enumerate}
    \item \textbf{Fostering cross-cultural collaboration:}
There is a valuable opportunity to create spaces for researchers to exchange insights on what aspects of this work are easily shared across contexts and what challenges remain. This could be facilitated by convening a community of researchers dedicated to cross-cultural studies, supported by the establishment of a Youth Advisory Panel. Such a panel would play a crucial role in interpreting the outputs of conversational agents across different countries and cultures, ensuring that local nuances and youth perspectives are meaningfully incorporated.
\item \textbf{Amplifying youth voices in AI design:}
There is a growing momentum for initiatives that bring youth voices directly into AI product development. Encouraging collaboration between young people, researchers, and developers is vital. Initiatives such as CERES or the International Centre for EdTech Impact, which bridge researchers and industry, offer valuable models for integrating youth perspectives to inform design, evaluation, and implementation processes.
\item \textbf{Informing policy and industry practice:}
Our work also highlights the importance of engaging policymakers on the issue of youth well-being, especially when co-design processes are involved. Raising awareness within both the policy and industry communities about the risks of output inaccuracies is essential. Research-informed design is critical to safeguarding young people’s well-being, ensuring that conversational agents contribute positively and do not cause unintended harm.
\end{enumerate}

\bibliographystyle{ACM-Reference-Format}
\bibliography{references}

\end{document}